\documentclass{ws-ijmpa}

\begin{document}

\markboth{H. S. Yang}
{Emergent Spacetime and The Cosmological Constant}

%
\catchline{}{}{}{}{}
%

\title{EMERGENT SPACETIME AND THE COSMOLOGICAL CONSTANT}

\author{HYUN SEOK YANG}

\address{School of Physics, Korea Institute for Advanced Study, Seoul 130-012, Korea \\
hsyang@kias.re.kr}

\maketitle

\begin{history}
\received{25 March 2008}
\end{history}

\begin{abstract}

We address issues on the origin of gravity and the dark energy (or the cosmological constant)
from the perspectives of emergent gravity. We discuss how the emergent gravity reveals a noble,
radically different picture about the origin of spacetime, which is crucial for a tenable solution
of the cosmological constant problem. In particular, the emergent gravity naturally explains
the dynamical origin of flat spacetime, which is absent in Einstein gravity.

\end{abstract}

\ccode{PACS numbers: 11.10.Nx, 98.80.Cq, 04.50.Kd}

\section{The Cosmological Constant and Emergent Spacetime}

We suggested in Ref. \refcite{hsy-cc} that the cosmological constant (CC) problem can be resolved
in a natural way if gravity emerges from a gauge theory in noncommutative (NC) spacetime.
Especially, it was shown that the emergent gravity formulated from a background independent theory reveals
a noble, radically different picture about the origin of spacetime where vacuum energy does not gravitate
but only fluctuations around the vacuum generate gravity.
An extreme example is that a flat spacetime emerges from uniform condensation of energy,
previously identified with the CC.

In this short contribution we will summarize an underlying spacetime picture emerging
from the background independent formulation of NC gauge theory.
Some technical details will be deferred to Refs. \refcite{hsy} and \refcite{hsy-review}.

In general relativity, gravity is described by metric fields of spacetime which is determined by
the distribution of matter and energy. That is, the spacetime geometry is not a rigid but a
dynamical object. But there is a blind point about the dynamical origin of spacetime
in general relativity; it says nothing about the dynamical origin of flat spacetime
since the flat spacetime is a geometry of special relativity rather than general relativity.
It turns out \cite{hsy-cc} that this nonchalance about the dynamical origin of flat spacetime
gives rise to the CC problem in general relativity.

However, it has been shown in Refs. \refcite{hsy}-\refcite{hsy-review}
that the emergent gravity from NC gauge theory naturally explains
the dynamical origin of flat spacetime, which is absent in Einstein gravity.
We want to emphasize that this spacetime picture is an inevitable consequence
provided one accepts the fact that gravity is an emergent phenomenon,
which nowadays becomes a new paradigm emerging from string theory.

If gravity is emergent from gauge fields, the general relativity implies that
spacetime should also be emergent from gauge field interactions \footnote{Here we suppose
that there exists a background independent theory where any spacetime structure
is not {\it a priori} assumed but defined from the theory.}.
If so, what kind of gauge field configurations corresponds to a flat spacetime ?
The most tenable answer is a uniform condensation of gauge fields in a vacuum, while
it appears as a vacuum energy in Einstein gravity, so causing the CC problem.
Therefore ``a solution" of the CC problem might be surprisingly
simple if and only if there exists a physically viable theory
(i.e., a background independent theory) to contain all the above properties.
The correspondence between NC field theory and gravity \cite{hsy,hsy-review} precisely
realizes the desired properties of emergent gravity and
so resolves the CC problem \cite{hsy-cc}.

A NC spacetime arises from a condensation of gauge fields in a vacuum:
\begin{equation} \label{nc-spacetime}
\langle B_{ab} \rangle_{\rm vac} = (\theta^{-1})_{ab} \;\; \Leftrightarrow
\;\; [y^a, y^b]_\star = i \theta^{ab} \;\; \Leftrightarrow \;\; [a_i, a_j^\dagger] = \delta_{ij}
\end{equation}
where $a, b=1, \dots, 2n$ and $i,j = 1, \cdots, n$. Every NC space can be represented as a theory of
operators in a Hilbert space ${\cal H}$, which consists of (NC) C*-algebra ${\cal A}_\theta$
like as a set of observables in quantum mechanics. A standard dynamical system
can be described in terms of vector fields as derivations of a certain (commutative) C*-algebra.
This concept of dynamics can be generalized to a NC space provided that
one describes NC dynamics in terms of derivations of NC C*-algebra.
This kind of vector fields in the NC space (\ref{nc-spacetime}) was explicitly constructed
in Ref. \refcite{hsy}, where it was shown that
the vector fields form an orthonormal frame and so define vielbeins of a gravitational metric.

Furthermore, if one has a representation of the Hilbert space ${\cal H}$,
e.g., a Fock space of harmonic oscillators for the space (\ref{nc-spacetime}),
any operator in ${\cal A}_\theta$ or any NC field can be represented as a matrix
whose size is determined by the dimension of ${\cal H}$. For the NC space (\ref{nc-spacetime}),
one gets $N \times N$ matrices in the $N \to \infty$ limit. In this sense,
the emergent geometry arising from the vector fields in the NC space (\ref{nc-spacetime})
can be understood as a dual geometry of large $N$ matrices in ${\cal H}$ according to
the large $N$ duality or AdS/CFT correspondence \cite{hsy,hsy-review}.

A looming point in emergent gravity is that a flat spacetime emerges from the vacuum
(\ref{nc-spacetime}) triggered by the Planck energy condensation. In other words,
the vacuum energy does not gravitate and so causes no CC problem \cite{hsy-cc}.
A flat spacetime is not free gratis but a result of
the maximum energy condensation in a vacuum.

\end{document}